\documentstyle[12pt]{article}
\begin{document}

\title {PRECISION TESTS OF THE MSSM.
\thanks{Supported in part by the Polish Committee for Scientific
        Research and European Union Under contract CHRX-CT92-0004.}}
\author{Piotr H. Chankowski \\
Institute of Theoretical Physics, Warsaw University\\
ul. Ho\.za 69, 00--681 Warsaw, Poland.\\
\\
Stefan Pokorski
\thanks{On leave of absence from
Institute of Theoretical Physics, Warsaw University}
\\
Max--Planck--Institute f\"ur Physik\\
Werner -- Heisenberg -- Institute \\
F\"ohringer Ring 6,
80805 Munich, Germany
}

\maketitle

\vspace{-12cm}
\begin{flushright}
{\bf MPI-PhT/95-49}\\
{\bf hep-ph 9505308}
\end{flushright}
\vspace{12cm}

\begin{abstract}
We present the results of a  global fit to the electroweak
observables in the MSSM in which, for the first time, all the
(relevant) low energy parameters of the model are treated as
independent variables. The best fit selects either very low or
very large values of ~$\tan\beta$ ~and chargino
(higgsino--like) and stop or/and the ~$CP-$odd Higgs boson are within the
reach of LEP 2. Moreover, the best fit gives ~
$\alpha_s(M_Z)=0.114\pm0.007$, ~which is lower than the
one obtained from the SM fits. The overall description of the
electroweak data is better than in the SM.
Those results follow mainly from the fact that in the MSSM one can increase
the value of ~
$R_b\equiv\Gamma_{Z^0\rightarrow\bar bb}/\Gamma_{Z^0\rightarrow hadrons}$ ~
{\it without} modyfying the SM predictions for other observables.
\end{abstract}

\newpage

Precision tests of the MSSM have been discussed by several groups
\cite{ELLIS1,ELLIS2,LANLUO,ABC,CW}. In particular, first global fit  to the
electroweak data within the MSSM parametrized in terms of few parameters
at the GUT scale is given in ref. \cite{ELLIS1}. In ref. \cite{ABC} the
so called ~$\epsilon_i$ ~parametrization is used and the r\^ole of light
superpartners is studied in some detail. Here we present the results
of a global fit to the electroweak observables with, for the first time,
all the (relevant) low energy parameters of the MSSM treated as
independent variables in the fit \cite{TAHOE}.
We believe the set of low energy parameters suggested by the
precision data may give an interesting hint on physics at the GUT scale.

Our strategy is analogous to the one often used for the SM:
in terms  of the best measured observables ~$G_F$, ~
$\alpha_{EM}$, ~$M_Z$ ~and the  less well known ~
$m_t$, ~$M_h$, ~$\alpha_s(M_Z)$ ~and a number of additional free parameters
in the MSSM  such as ~$\tan\beta$, ~$M_A$, ~soft SUSY breaking scalar masses,
trilinear couplings etc. we calculate in the MSSM the observables ~$M_W$, ~
all partial widths of ~$Z^0$ ~and all asymetries at the ~$Z^0$ ~pole. This
calculation is performed in the on--shell renormalization scheme
\cite{HOL,HOLL} and with
the same precision as the analogous calculation in the SM, i.e. we include
all supersymmetric oblique and process dependent one--loop corrections
\cite{MY_H,MY_DR}, and also the leading higher order effects.

Similar programme has often been discussed in the context of the SM
\cite{ELLIS3,ALBA,LANER,ELLIS2,EWG,MY_SMH} with the parameters ~
$m_t$, ~$\alpha_s(M_Z)$ ~and ~$M_h$ ~(or some of them) to be determined
by a fit to the data.
Let us first review the results in the SM with the emphasis on those
features which are relevant for the supersymmetric extension.

As the experimental
input for the observables ~$1-M^2_W/M^2_Z$, ~$\Gamma_Z$, ~
$\sigma_{h}$, ~${\cal A}_e$,  ~${\cal A}_{\tau}$, ~
$\sin^2\theta^{lept}_{eff}<Q_{FB}>$, ~${\cal R}_l$, ~$A_{FB}^{0,l}$, ~
{}~$R_b$, ~$R_c$, ~$A_{FB}^{0,b}$, ~$A_{FB}^{0,c}$, ~
(i.e. their experimental values,
errors and correlation matrices) used in the fits we take the
Spring 95 data summarized in ref. \cite{EWG}. For the top quark
mass (which we include in the fit) we use the weighted average
of the CDF and D0 result ~$m_t=(181\pm12)$ GeV \cite{CDF_MT}.
For ~$M_W$ ~we use the results of ref. \cite{MW}: ~$M_W = 80.33\pm0.17$
GeV. ~(This is the average
value of the UA2 measurement and the new measurement reported by the
CDF \cite{MW}. The D0 collaboration has not published the results
of their new analysis yet.)

The Left -- Right asymmetry measured by the SLD is ~$A_{LR}=0.1551\pm0.0040$ ~
\cite{SLD} and it is included in the fit.
For the value of ~$\Delta\alpha^{hadr}_{EM}$ ~we use the result of the
recent re--analysis in  ref. \cite{JEG}: ~
$\Delta\alpha^{hadr}_{EM}=0.0280\pm0.0007$ ~with the error propagating
in the fit
\footnote{The result reported in the ref. \cite{SWA} has
been recently updated  \cite{PES} and are now closer to the results
reported in \cite{JEG} and those in ref \cite{ZEP} are based on more
theoretical assumptions \cite{BRUSS}.}.

In Table 1 ~we present the results of our global fit in the SM
\footnote{Those results agree very well with another  recent fit \cite{EL}.}
(for the sake of later discussion
in version B we include in the fit the low energy measurement of ~
$\alpha_s$ ~\cite{LOW}: ~$\alpha_s(M_Z)=0.112\pm0.005$).

\begin{center}
{\bf Table 1.} Results of a fit in the SM. All masses in GeV.
\vskip 0.2cm
\begin{tabular}{||l||l|l|l|l|l|l|l|l||} \hline
 fit             &
$m_t$      &
$\Delta m_t$   &
$M_h$           &
$\Delta M_h$    &
$\alpha_s(M_Z)$ &
$\Delta\alpha_s(M_Z)$ &
$\chi^2$         &
d.o.f            \\ \hline
A&171.3&$^{+11.5}_{-9.7}$&66&$^{+117}_{-45}$&0.123&0.005&12.6&12\\ \hline
B&172.0&$^{+10.5}_{-9.3}$&59&$^{+96}_{-37}$&0.120&0.005&15.5&13\\ \hline
\end{tabular}
\end{center}

The fitted value for ~$M_h$ ~ results in a very transparent
way from a combination of effects which can be organized into the
following two--step description \cite{MY_SMH}. A fit to ~$M_W$ ~
and to all measured electroweak observables but ~
$R_b\equiv\Gamma_{Z^0\rightarrow\bar bb}/\Gamma_{Z^0\rightarrow hadrons}$ ~
gives ~$\chi^2$ ~values which are almost
{\it independent} of the value of the top quark mass ~$m_t$ ~
in the broad range (150--200 GeV) and with the best value of ~
$\log M_h$ ~which is almost linearly correlated with ~$m_t$. ~
This is shown in Fig.1. ~
The ~$m_t-M_h$ ~correlation is the most solid result of the fits
which does not depend on whether ~$R_b$ ~and/or ~$m_t$ ~measurement of
the CDF and D0 \cite{CDF_MT} are included in the fits. It points toward
relatively light Higgs boson for ~$m_t$ ~in the range ~$(170 - 190)$ GeV. ~

A visible ~$\chi^2$ ~dependence as a function of ~$m_t$ ~
and, therefore, indirect (by constraining ~$m_t$) ~relevant
overall limit on the Higgs mass ~$M_h$ ~is introduced by the
results for ~$R_b$ ~and ~$m_t$. ~This is also  clearly seen in Fig.1.

Another point of recent interest is the value of ~$\alpha_s(M_Z)$ ~
obtained from the electroweak fits. It is somewhat larger then the
value  obtained from low energy data \cite{LOW}.
It is interesting to repeat the SM fit with the
low energy measurement ~$\alpha_s(M_Z)=0.112\pm0.005$ ~
included in the fit.
\footnote{One can argue that the determination
of $\alpha_s(M_Z)$ based on the deep inelastic (Euclidean) analysis
is more precise than from the experiments in the Minkowskean region
(jet physics, $\tau$ decays). Low value of $\alpha_s(M_Z)$ is also
consistent with lattice calculation and has some theoretical support
(for review of all those points see M.Shifman, ref.[25]). So, with proper
attention to the unsettled controversy and to the fact that jet physics
and $\tau$ decays give larger values, we are going to explore the
assumption that the low energy determination of $\alpha_s(M_Z)$
is the correct one.}
Those results are also shown in in Table 1 (case B).
The parameters of the fit remain almost unaltered but the overall ~$\chi^2$ ~
is larger by ~$\sim 3$.

Finally we can interpret the SM fits as the MSSM fits with all
superpartners heavy enough to be decoupled. Supersymmetry then just
provides a rationale for a light Higgs boson: $M_h \sim {\cal O}$(100 GeV).
and we can expect that the MSSM with heavy enough superpartners gives
as good a fit to the precision electroweak data as the SM , with ~
$m_t\sim170 - 180$ GeV ~(depending slightly on the value
of ~$\tan\beta$).

This is seen in Fig.2 where we show the ~$\chi^2$ ~values in the MSSM
in both versions, A and B,
with the proper dependence of ~$M_h$ ~on ~$m_t$, ~$\tan\beta$ ~and
SUSY parameters included \cite{ZWIR}, with  fitted  ~$m_t$ ~and ~
$\alpha_s(M_Z)$ ~and with all SUSY mass parameters fixed at 500 GeV.
The ~$\chi^2$ ~values in the minima are very closed to  the SM  three
parameter ~($m_t$, ~$M_h$, ~$\alpha_s(M_Z)$) ~
fit. ~The only difference is in the ~$m_t$ ~dependence of ~$\chi^2$: ~
the minimum in the MSSM fit is for slightly larger ~$m_t$ ~and
simultaneously the upper bound on ~$m_t$ ~is more stringent ~
($m_t < 188$ GeV ~at 95\% C.L.).
This is easy to understand as due to the very constrained ~$M_h$ ~
in the model and to ~$M_h - m_t$ ~correlation needed to fit the data
\footnote{Scanning over ~$M_A$ ~does not change this result as
significantly lower values of ~$M_A$ ~are excluded by ~
$b\rightarrow s\gamma$ ~and/or worsen the fit due to negative
contribution to ~$R_b$.}.

Although the SM fit and the MSSM fit with heavy superpartners are
globally good, it has been noticed that they cannot properly account
for the measured value of ~$R_b$ ~which remains almost ~$3\sigma$ ~
higher than the theoretical prediction. Moreover it is well known that
new physics in ~$\Gamma_{Z^0\rightarrow\bar bb}$ ~and therefore
additional contribution to the total hadronic width of the ~$Z^0$ ~
boson would lower
\footnote{In the electroweak fits the value of ~$\alpha_s(M_Z)$ ~is very
precisely determined by strong corrections to the total hadronic ~$Z^0$ ~
width. This quantity is calculated with high precision (up to ~
${\cal O}(\alpha_s^3)$) ~and the experimental error is also very small: ~
$\Gamma_h = 1744.8\pm3.0$ \cite{BRUSS}.}
the fitted value of ~
$\alpha_s(M_Z)$ ~\cite{BV,LANER}, in better agreement with its
determination from low energy data \cite{LOW}.

Thus it is conceivable that the measurement of ~$R_b$ ~is not a
statistical fluctuation but an evidence for new physics and it is very
interesting to perform a global fit to the electroweak observables in
the MSSM with supersymmetric masses kept as free parameters.
In particular we can ask the following two questions \cite{TAHOE}:

\noindent a) can we improve ~$R_b$ ~without destroying the excellent
fit to the other observables?

\noindent b) if we achieve this goal, what are the predictions for
sparticle masses?

We begin with a brief overview of the SUSY corrections to the electroweak
observables. Although
the MSSM contains many free SUSY parameters several of them are irrelevant.
Here we list the fitted parameters. These are: ~
$m_t$, ~$\alpha_s(M_Z)$, ~$\tan\beta$, ~$M_A$, ~$\mu$, ~$M_{g2}$ ~(we use
the relation ~$M_{g1} = (5/3)\tan^2\theta M_{g2}$; it is of little
importance for the results of the fit but fixes the parameters of the
lightest supersymmetric particle LSP), ~$m_{\tilde q_L}$ ~
(the soft mass term for the third generation left -handed squarks),
$m_{\tilde b_R}$, ~$m_{\tilde t_R}$, ~$A_b$, ~$A_t$ ~and ~
$m_{\tilde l_L}$ ~(a common soft mass parameter
for all left handed sleptons). ~
The gluino mass,
the first two generation squark masses and the right handed slepton masses
are irrelevant for the fit and are always kept heavy.
Furthemore, there are remarkable regularities
in SUSY corrections \cite{TAHOE}. Following our strategy
of calculating all electroweak observables in terms of ~$G_F$, ~$M_Z$ ~and ~
$\alpha_{EM}(0)$ ~one can establish the following ``theorems'' for the
predictions in the MSSM:

\vskip 0.2cm
\noindent 1.) $(M_W)^{MSSM} \geq (M_W)^{SM}$.
As explained in ref. \cite{MY_DR}, its origin lies mainly
in additional sources of
the custodial ~$SU_V(2)$ ~violation in the squark and slepton left-handed
mass matrix elements (we denote them with capital letters e.g. ~
$M_{\tilde t_L}^2 = m_{\tilde q_L}^2 + m_t^2 + t_{\beta}(M^2_Z - 4M^2_W)$ ~
and similarly for the other squarks and sleptons):
\begin{eqnarray}
M^2_{\tilde l_L} - M^2_{\tilde\nu} = t_{\beta} M^2_W, ~~~~~
M^2_{\tilde t_L} - M^2_{\tilde b_L} = m^2_t - m^2_b - t_{\beta}M_W^2
\label{eqn:m_split}
\end{eqnarray}
where ~$t_{\beta}\equiv(\tan^2\beta-1)/(\tan^2\beta+1)$, ~which contribute
to ~$\Delta\rho$ ~with the same sign as the ~$t - b$ ~mass splitting.
It should be stressed that the
supersymmetric prediction for ~$M_W$ ~is merely sensitive
to ~$m_{\tilde l_L}$ ~and ~$m_{\tilde q_L}$ ~which determine the magnitude
of the splitting in eq.(1) relative to the masses ~$M_{\tilde t_L}$ ~etc.
The dependence
on the right--handed sfermion masses
enters only through the left--right mixing. This also means that the
predicted $M_W$ is almost insensitive to the masses of squarks of the first
two generations: in their left--handed components there is no source of large ~
$SU_V(2)$ ~violation. Also, the predictions for ~$M_W$ ~are rather weakly
dependent on the chargino and neutralino masses ~$m_{C^\pm}$, ~$m_{N^0}$ ~
and the Higgs sector parameters
\footnote{This is due to generically weak $SU_V(2)$ breaking effects in
          these sectors.}.
\vskip 0.2cm
\noindent 2.) Another effect of supersymmetric corrections is that ~
$(\sin^2\theta^{\it l})^{MSSM} \leq (\sin^2\theta^{\it l})^{SM}$ ~where
$\sin^2\theta^{\it l}$ ~can be determined from the
on--resonance forward--backward asymmetries
\begin{eqnarray}
A_{FB}^{0~\it l} = {3\over4}{\cal A}_e{\cal A}_{\it l}
{}~~{\rm where} ~~
{\cal  A}_f = {2x_f\over 1+ x^2_f}
\end{eqnarray}
with ~$x_f = 1 - 4|Q_f|\sin^2\theta^f$.
In general, in the on--shell renormalization scheme and with the loop
corrections included we get:
\begin{eqnarray}
\sin^2\theta^{\it l}=
\left(1 - {M^2_W\over M^2_Z}\right)\kappa_{UN}(1+ \Delta\kappa_{NON})
\label{eqn:sin_for}
\end{eqnarray}
where $\kappa_{UN}$ contains universal ``oblique'' corrections and
$\Delta\kappa_{NON}$ -- genuine (nonuniversal) vertex corrections which
are in this
case negligibly small. By using explicit form of $\kappa_{UN}$ ~one can derive
the following relation:
\begin{eqnarray}
(\sin^2\theta^{\it l})^{MSSM} = (\sin^2\theta^{\it l})^{SM}\times
\left[1 - {c^2_W\over c^2_W - s^2_W}
\left(\Delta\rho\right)^{SUSY} + ...\right]
\end{eqnarray}
We see therefore, that the supersymmetric predictions for ~
$\sin^2\theta^{\it l}$ ~are correlated with the predictions for ~$M_W$ ~
through the value of ~$\Delta\rho$ ~
and they are sensitive to the same supersymmetric parameters.
\vskip 0.2cm

3.) Similarly, the asymmetries in the quark channel are given by the product
\begin{eqnarray}
A_{FB}^{0~\it q} = {3\over4}{\cal A}_e{\cal A}_q
\end{eqnarray}
If ~
$(\sin^2\theta^{\it l})^{MSSM}=(\sin^2\theta^{\it l})^{SM} - \varepsilon$ ~
and ~$(\sin^2\theta^b)^{MSSM}=(\sin^2\theta^b)^{SM} - \delta$ ~
then it is easy to show that
\begin{eqnarray}
(A_{FB}^{0~\it b})^{MSSM} =(A_{FB}^{0~\it b})^{SM} \times
\left(1 + {\varepsilon\over1-4\sin^2\theta^{\it l}} + 0.2\delta\right)
\end{eqnarray}
Thus, supersymmetric corrections to ~$A_{FB}^{0~\it b}$ ~are essentially
determined
by the corrections to ~$\sin^2\theta^{\it l}$ ~and give
the third ``theorem'': ~
$(A_{FB}^{0~\it b})^{MSSM}\geq (A_{FB}^{0~\it b})^{SM}$. ~

At this point it is important to observe that the trends in the MSSM
expressed by the above three theorems can only make the comparison of the
MSSM predictions with the data worse than in the SM (as for ~$m_t > 170$
GeV ~they go against
the trend of the data!). Thus we can expect to get lower limits
on the left--handed squark and slepton masses, which are the parameters
most relevant for the observables ~$M_W$, ~$\sin^2\theta^{\it l}$ ~and ~
$A_{FB}^{0~\it b}$ ~(of course, for large enough masses, we recover the SM
predictions). These limits are amplified by the dependence of ~$\Gamma_Z$ ~
on ~$m_{\tilde q_L}$ ~and ~$m_{\tilde {\it l}_L}$. ~As discussed above, the
sensitivity of the each one of those  observables to the remaining parameters
is weak but may become nonnegligible in the global fit.
The most important and interesting is the dependence on the chargino
mass and its composition. One can see that a light higgsino (and only
higgsino) does not worsen the predictions for ~
$M_W$, ~$\sin^2\theta^{\it l}$, ~$A_{FB}^{0~\it b}$ ~and, for a heavy top
quark and light Higgs boson it can significantly improve the fit to ~
$\Gamma_Z$ ~\cite{BARB} due to the ~$Z^0$-wave function renormalization
effect which acts similarly to a heavier Higgs boson for heavier $t$ quark
(i.e. its contribution makes ~$\Gamma_Z$ ~smaller).\footnote{However,
this effect can be masked by the additional contribution to the width from
$Z^0\rightarrow N_i^0N_j^0$~(see later).}

Finally, let us discuss the
corrections to the ~$Z^0\rightarrow\overline b b$ ~
vertex which contribute to the observable ~$R_b$.

In the MSSM there are three types of important corrections to the vertex ~
$Z^0\overline b b$ ~\cite{BF}:

\noindent a) charged and neutral Higgs boson exchange \cite{JR}; for low ~
$\tan\beta$ ~
and light ~$CP-$odd Higgs boson ~$A^0$ ~this contribution is negative (the ~
$\Gamma_{Z\rightarrow\bar bb}$ ~is decreasing below its SM value) whereas
for very light ~$A^0$ ~(50 -- 80 GeV) and very large values of the ~
$\tan\beta$ ~
($\sim50$) ~the interplay of charged and neutral Higgs bosons is strongly
positive;

\noindent b) chargino -- stop loops; for heavy top quark and small ~
$\tan\beta$ ~(i.e. for a given ~$m_t$ ~-- maximally large top quark
Yukawa coupling)
they can contribute significantly (and positively) for light chargino (if
higgsino-like) and light right-handed top squark (this follows from the
Yukawa chargino--stop--bottom coupling); in the case of large $\tan\beta$
this contribution is smaller than for small ~$\tan\beta$ ~but the total
contribution to the ~$Z^0\bar b b$ ~vertex can be amplified by

\noindent c) neutralino -- sbottom (if light) loops.

Thus, in the MSSM the value of ~$R_b$ ~can in principle be significantly
larger than in the SM for very low or very large values of ~$\tan\beta$, ~
light (higgsino-like) chargino, and ~$\tilde t_R$ ~and/or very light ~
$A^0$ ~(for large ~$\tan\beta$) ~\cite{KKW}. It is insensitive to ~
$m_{\tilde q_L}$ ~and ~$m_{\tilde {\it l}_L}$.

In summary, in MSSM the electroweak observables exhibit
certain ``decoupling'': all of them but ~$R_b$ ~are sensitive mainly
to the left--handed slepton and the third generation squark masses and
depend weakly on the right--handed squark masses, gaugino
and Higgs sectors; on the contrary, ~$R_b$ ~depends strongly just
on the latter set of variables and very weakly on the former. We can
then indeed expect to increase the value of ~$R_b$ ~ without destroying
the perfect fit of the SM to the other observables. However,
chargino, right--handed stop and charged Higgs boson masses also
are crucial variables for the decay ~$b\rightarrow s\gamma$ ~ and this
constraint has to be included (there is a sizable  uncertainty
in the theoretical prediction for ~
$BR(b\rightarrow s\gamma)$ ~\cite{BSG_BG,BMMP} which is taken into
account in this paper
\footnote{It is interesting to observe that the uncertainty in the
renormalization scale ~$\mu$ ~in the standard formula ~$C^{eff}_7(\mu)=
\eta_7(\mu)C_{\gamma}(M_W)+\eta_8(\mu)C_g(M_W)+\eta_2(\mu)C_2(M_W)$ ~
(where ~$\eta_i(\mu)$ ~are model independent QCD corrections)
can give much larger uncertainty (up to factor two) in the full
amplitude in the MSSM compared to the SM. This is due to the fact that,
with supersymmetric contributions, the first two terms in ~$C^{eff}_7$ ~
can be positive whereas in the SM all three are negative;  in the latter
case an increase by about 80\% of ~$\eta_2(\mu)$ ~when ~$\mu$ ~changes from ~
$2m_b$ ~to ~$m_b/2$ ~is partially compensated by decreasing
$\eta_7$ ~and ~$\eta_8$; ~with positive ~$C_{\gamma}$, ~$C_g$ ~and
negative ~$C_2$ ~both effects can add up.}).

Of course the obvious constraint for our fits are the present
experimental lower bounds for superpartner masses. For
chargino and stop we take them to be  47 GeV.
In addition, in the parameter space which gives light higgsino--like
charginos also neutralinos are higgsino--like and therefore the
contribution of ~
$Z^0\rightarrow N^0_iN^0_j$ ~to the total ~$Z^0$ ~width is important
in the fit (we also impose the constraint that ~$N^0_1$ ~ is the LSP).
Another important constraint follows from non--standard top quark decays
such as ~$t\rightarrow\tilde t N^0_i$ ~for a light stop ~$\tilde t$. ~
The discovery of the top quark in Fermilab through standard decay modes
puts upper bound on the ~$BR$ ~for non-standard top decays which is of the
order of 50\% \cite{QUIGG}.

Let us present some quantitative results.
We have fitted the value of ~$\alpha_s$, ~$\tan\beta$, ~$m_t$ ~and SUSY
parameters listed earlier in two versions A and B (without and with the
low energy value ~$\alpha_s(M_Z)=0.112\pm0.005$ ~in the fit).
The dependence of the ~$\chi^2$ ~on ~$\tan\beta$ ~for several values of ~
$m_t$ ~(and scanned over the other parameters) is shown in ~Fig.3. ~The best
fit is obtained in two regions of very low (close to the  quasi--IR
fixed point for a given top quark mass, see e.g. \cite{CPW} and
references therein) and very large ~$(\sim m_t/m_b)$ ~$\tan\beta$ ~values
(for early discussion of large ~$\tan\beta$ ~region see \cite{OP}).
The results are summarized in Table 2.
\newpage
\begin{center}
{\bf Table 2.} Results of a fit in the MSSM.
\vskip 0.2cm
\begin{tabular}{||l|l||l|l|l|l||} \hline
 fit                       &
$\tan\beta$                &
$m_t$                      &
$\alpha_s(M_Z)$            &
$\chi^2$                   &
$R_b$                      \\ \hline\hline
A& $IR$&$178^{+5}_{-8}$&$0.116^{+0.006}_{-0.004}$&10.3&0.218\\ \hline
B& $IR$&$177^{+4}_{-6}$&$0.114^{+0.004}_{-0.003}$&10.6&0.218\\ \hline\hline
A& $m_t/m_b$&$172^{+8}_{-7}$&$0.114\pm0.005$&10.2&0.219\\ \hline
B& $m_t/m_b$&$174^{+6}_{-7.3}$&$0.113\pm0.004$&10.2&0.219\\ \hline\hline
\end{tabular}
\end{center}

We recall (see Fig. 2) that in the fit with all superpartners heavy the
best $\chi^2$  values read:  ~$\chi^2=13(16)$,
for ~$\tan\beta=1.4$ ~and ~
 ~$\chi^2=13.3(16)$,
for ~$\tan\beta=50$ ~for fits without (with) low
energy value for ~$\alpha_s$ ~included. We observe that in version A (B)
the best values of ~$\chi^2$ ~are by 3(6) lower than in the corresponding
fits with all superpartners heavy. Clearly, in version A this improvement
is mainly due to higher values of ~$R_b$ ~whereas in version B also to
the fact that the fitted values of ~$\alpha_s(M_Z)$ ~are lower than in
the fit with heavy superpartners and much closer to the low energy value ~
$\alpha_s(M_Z)=0.112\pm0.005$ ~which in version B is included in the fit.
It is well known that additional contributions to ~
$\Gamma_{Z\rightarrow\bar bb}$ ~lower the fitted value of ~
$\alpha_s(M_Z)$ ~\cite{BV,LANER} and this effect is indeed observed in
our fits.

In Fig. 4a we show ~$\chi^2$ ~values for versions A of the fit as
a function of ~$\alpha_s(M_Z)$ ~for different values of ~$\tan\beta$ ~
and in Fig. 4b the global dependence of ~$\chi^2$ ~on ~$\alpha_s(M_Z)$, ~
with the best fit for ~$\alpha_s(M_Z)=0.114\pm0.007$. ~
The global dependence of ~$\chi^2$ ~on ~$m_t$ ~is shown in Fig. 5.
We get ~$m_t=175^{+7}_{-9}$ GeV ~and ~$176^{+4}_{-8}$ GeV ~for the
A and B versions of the fit respectively.

Increase of ~
$\Gamma_{Z\rightarrow\bar bb}$ ~ requires light stop and
chargino (for low values of ~$\tan\beta$) ~or light ~$CP-$odd scalar
and/or chargino and stop for large ~$\tan\beta$ ~and it is bounded
from above by the experimental lower limits on the masses of those
particles. A light and dominantly right-handed stop
\begin{equation}
M^2_{\tilde t_1} = M^2_{t_R}  - 2\theta^2 M^2_{\tilde t_L},
{}~~~~~~~|\theta| = |\frac{A_tm_t}{M^2_{\tilde t_L}}| << 1
\label{eqn:stop}
\end{equation}
is obtained
for ~$M_{\tilde t_L} >>M_{\tilde t_R}$ ~and large L--R mixing term ~
$A_t/M_{\tilde t_L}\sim$ \\
$\sqrt{(M^2_{\tilde t_R}-M^2_{\tilde t_1})/
2m^2_t}$ ~(we recall that in our notation capital letters denote
the full diagonal entries in the sfermion mass matrix).
Our fits give upper bounds on the light stop,
chargino and CP--odd Higgs boson masses. In version A, when $\alpha_s$
runs free and is fitted only to the electroweak data, the best fit
is better than the corresponding fit with all superpartners heavy
by only ~$\Delta\chi^2\sim 3$ ~(but then ~$\alpha_s(M_Z)=0.123$). ~
So, we obtain strong upper bounds at ~1$\sigma$ ~level but no 95\% C.L. limits.
They are shown in Fig.6a and 7a for the stop and chargino masses in
the low ~$\tan\beta$ ~region and for the pseudoscalar and chargino masses
in the large $\tan\beta$ region, respectively. Stronger bounds are obtained
in version B of the fits, i.e. with the low energy value of ~$\alpha_s(M_Z)$ ~
included in the fits. They are shown in the same Figures. The strongest
bounds are obtained when ~$\alpha_s(M_Z)$ ~is fixed to its best fit value
and they are shown in Fig.6b and 7b. The dependence of the strength
of the bounds on the way we treat $\alpha_s$ in our fits is quite obvious
from the earlier discussion of the depth of the minima in ~$\chi^2$. ~
It is interesting to note the structure of the bounds in Figures 7:
although the ~2$\sigma$ ~bounds never constrain the pseudoscalar and the
chargino masses simultaneously, one of them remains always light.

Finally in Fig.8 we show the lower ~1 ~and ~2$\sigma$ ~limits on
left--handed sbottom and left--handed slepton masses for different ~
$m_t$ ~and ~$\tan\beta$. ~

The ~$BR(b\rightarrow s\gamma)$ ~has been calculated for each point
of the fit with the theoretical uncertainty included according to
the ref. \cite{BMMP}. Only the points with ~$BR(b\rightarrow s\gamma)$ ~
within ~$2\sigma$ ~of the experimental result have been retained.
In Fig. 9 ~we present the scatter plots which illustrate the role
of this constraint in the large ~$\tan\beta$ ~region. In Fig. 9a ~
the points with ~$\Delta\chi^2 < 4$ ~are plotted in the ~
$(R_b,~BR(b\rightarrow s\gamma))$ ~plane and in Fig. 9b we show ~
$\chi^2$ ~versus ~$BR(b\rightarrow s\gamma)$. ~One can see that the
requirement of acceptable ~$BR(b\rightarrow s\gamma)$ ~rejects part of
the points with best ~$R_b$ ~and ~$\chi^2$ ~but is
consistent with a large number of such points.

In general, one obtains acceptable ~$BR(b\rightarrow s\gamma)$ ~due to
cancellations between ~$W^{\pm}$, ~$H^{\pm}$ ~and ~$C^{\pm}$ ~and ~
$\tilde t_R$ ~loops. The net impact on the allowed parameters space depends
quite strongly on the values of ~$m_t$ ~and ~$\alpha_s(M_Z)$. ~
In the approximation of refs. \cite{BSG_BG} and in the limit of
pure higgsino--like chargino and very heavy second stop and second chargino
the amplitude for ~$b\rightarrow s\gamma$ ~(before QCD corrections) reads:
\begin{eqnarray}
A_{b\rightarrow s\gamma} = \sum_{i=\gamma,g}(A^i_W + A^i_{H^+} + A^i_C)
\end{eqnarray}
\begin{eqnarray}
A^i_W + A^i_{H^+} =
{3\over2}{m_t^2\over M_W^2} ~f^{(1)}_i\left({m^2_t\over M^2_W}\right)
+ {1\over2}{m_t^2\over M_{H^+}^2} ~f^{(2)}_i
\left({m^2_t\over M^2_{H^+}}\right)
\end{eqnarray}
\begin{eqnarray}
A^i_C = -\lambda {m^2_t\over m_{C_1}^2}\left[
f^{(1)}_i\left({M^2_{\tilde t_1}\over m^2_{C_1}}\right)
+ \tan\beta {m_{C_1}A_t\over M^2_{\tilde t_L}} ~
f^{(3)}_i\left({M^2_{\tilde t_1}\over m^2_{C_1}}\right)\right]
\end{eqnarray}
The functions ~$f^{(k)}_{g,\gamma}$ ~k=1,2,3 ~are  defined in \cite{BSG_BG}
and they all take negative values,
the factor ~$\lambda\approx1$ ~for small values of ~$\tan\beta$ ~and ~
$\lambda\approx 1/2$ ~for large values of ~$\tan\beta$. ~
We see that e.g. for  large ~$\tan\beta$, ~with small ~$M_A$, ~
$m_C$ ~and ~$M_{\tilde t_1}$ (as needed for the largest $R_b$)
acceptable ~$BR(b\rightarrow s\gamma)$ ~requires cancellation between ~
$(A^i_W+A^i_{H^+})$ ~and ~$A^i_C$ ~(which has to be positive) and
correlates those masses and the Left--Right
mixing angle ~$\theta_{\tilde t}\approx -m_tA_t/M^2_{\tilde t_L}$ ~
(up to the experimental and theoretical uncertainties in the ~$BR(b\rightarrow
s\gamma)$. ~
Although we have to cancel large ~$H^+$ ~contribution,
due to the large ~$\tan\beta$ ~value the mixing angle
which is needed remains small in agreement with the angle ~$\theta$ ~
in eq. (\ref{eqn:stop}).
The first equation in (\ref{eqn:stop}) can be satisfied by a
proper adjustement of the parameter ~$M_{\tilde t_R}$. ~This
explains the pattern seen in Fig.9.

In summary, the MSSM fit to the electroweak observables is very good.~
This is mainly due to higher than in the SM values of ~$R_b$, ~
without destroying the agreement in the other observables.
Moreover the best fit gives ~$\alpha_s(M_Z)=0.114\pm0.007$, ~a
value which is lower than the one obtained from
the SM fits and in agreement with the low energy data. Low value
of ~$\alpha_s(M_Z)$ ~is correlated with the presence of the additional
contribution to the ~$\Gamma_{Z\rightarrow\bar bb}$ ~\cite{BV,LANER}.
The best fit selects very particular regions of the parameters space:
either very low or very large values of ~$\tan\beta$ ~and
small higgsino--like chargino and right--handed stop masses for low $\tan\beta$
or/and  the ~$CP-$odd ~Higgs boson mass for large $\tan\beta$.
For the best value of ~$\alpha_s(M_Z)$ ~or with the low energy
measurement of ~$\alpha_s(M_Z)$ ~included in the fit we obtain strong
upper bounds at 95\% C.L.
on the masses of these particles and predict that they are
within the reach of LEP2.

Furthemore, the parameter space selected by the fit is interesting from the
theoretical point of view. Low and large values of ~$\tan\beta$ ~
are theoretically most appealing \cite{CPW,OP}. The hierarchy ~
$M_{\tilde t_L} > M_{\tilde t_R}$ ~which is necessary for a good fit
in the low ~$\tan\beta$ ~region can be viewed as a natural effect of the
top quark Yukawa coupling in the renormalization group running from the
GUT scale. For a good fit in the large ~$\tan\beta$ ~region this hierarchy
is less pronounced, in agreement with ~$Y_t\approx Y_b$. ~
Finally, the hierarchy ~$\mu << M_{g2}$ ~(i.e. higgsino--like lightest
neutralino and chargino) is inconsistent with the mechanism of
radiative electroweak symmetry breaking and universal boundary conditions
for the scalar masses at the GUT scale in the minimal supergravity
model. However, it is predicted in models with certain pattern of
non--universal boundary conditions \cite{OP2}.

\vskip 0.5cm
Qualitatively similar conclusions for small ~$\tan\beta$ ~have been also
reached in a recent paper \cite{KANE}.
For large ~$\tan\beta$, ~value of ~$\alpha_s(M_Z)$ ~similar
to ours has been obtained recently also in \cite{SOLA2}.
\vskip 0.5cm

{\bf Acknowledgments:} P.Ch. would like to thank
 Max--Planck--Institut f\"ur
Physik for warm hospitality during his stay in Munich where
part of this work was done.

{\bf Note added.} New electroweak data have been presented at the
International Europhysics Conference on High Energy Physics
(Brussels, 27 July -- 2 August, 1995). The main change are the values
of ~$R_b$ ~and ~$R_c$: ~$R_b=0.2219\pm0.0017$, ~$R_c=0.1543\pm0.0074$. ~
Since identification of the ~$c$ ~quarks is more difficult than of the ~
$b$ ~quarks, the experimental groups also present the value of ~
$R_b = 0.2206\pm0.0016$ ~obtained under the assumption that ~$R_c$ ~is
fixed to its SM value ~$R_c=0.172$. ~The results of the present paper
remain unchanged if we adopt the latter value of ~$R_b$ ~and disregard
the new value of ~$R_c$ ~as unreliable. The MSSM cannot explain any
significant departure of ~$R_c$ ~from the SM prediction ~\cite{SOLA3}.

\newpage

\noindent {\bf FIGURE CAPTIONS}
\vskip 0.5cm

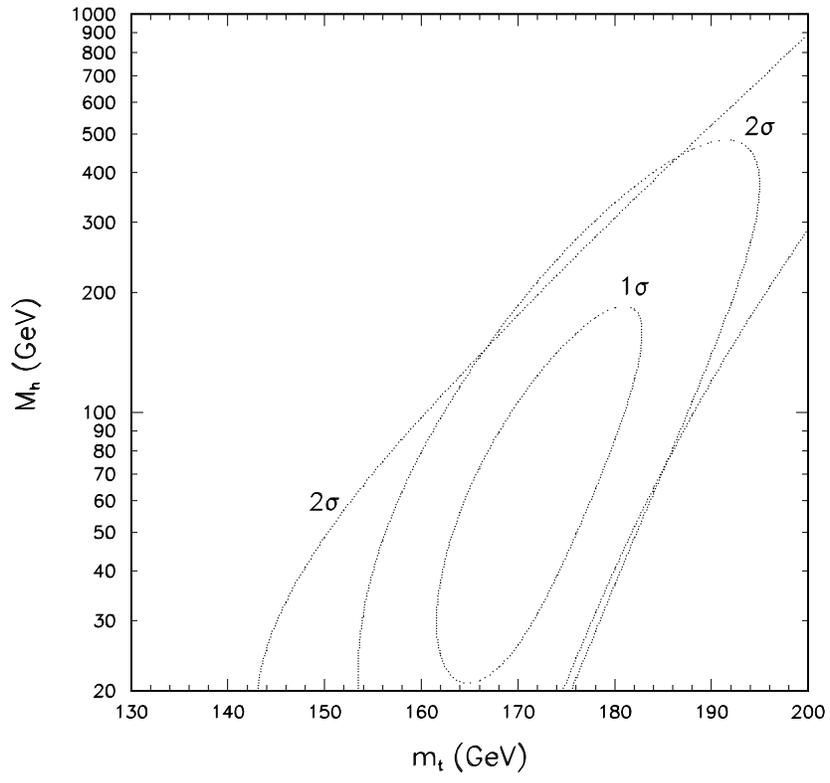
\noindent {\bf Figure 1.}
Limits in the ~$(m_t, M_h)$ ~plane in the SM.
Unclosed lines show the ~2$\sigma$ ~limits from the fit
without the ~$R_b$ ~and ~$m_t$ ~measurements included.
Ellipses show the ~1$\sigma$ ~and ~2$\sigma$ ~limits from
the fit with the ~$R_b$ ~and ~$m_t$ ~measurements included.
\vskip 0.3cm

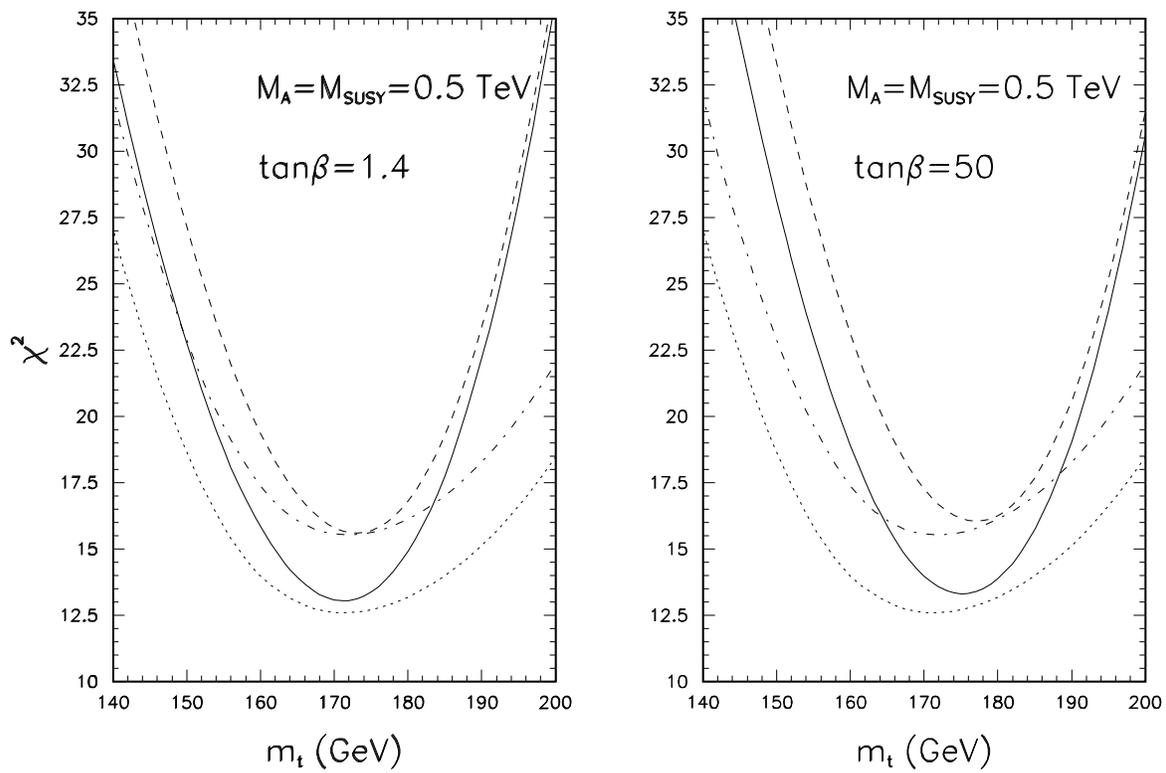
\noindent {\bf Figure 2.}
$\chi^2$ ~as a function of ~$m_t$ ~
for two values of ~$\tan\beta$ ~in the MSSM with
heavy superparticles.
Solid (dashed) lines show  ~$\chi^2$ ~without (with) the
low energy measurement of ~$\alpha_s$ ~
($\alpha_s(M_Z)=0.112\pm0.005$) ~included in the fit (versions
A and B of the fit respectively). For comparison, the corresponding
fits in the Standard Model are shown with dotted (dash--dotted)
lines.
\vskip 0.3cm

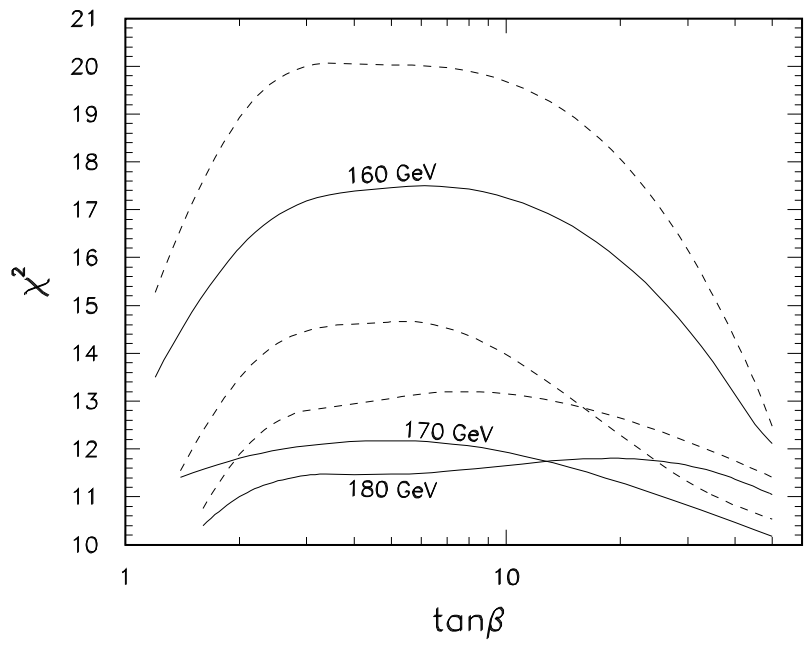
\noindent {\bf Figure 3.}
Dependence of ~$\chi^2$ ~on ~$\tan\beta$ ~for different
values of ~$m_t$. ~Solid and dashed lines correspond to versions
A and B of the fit respectively.
\vskip 0.3cm

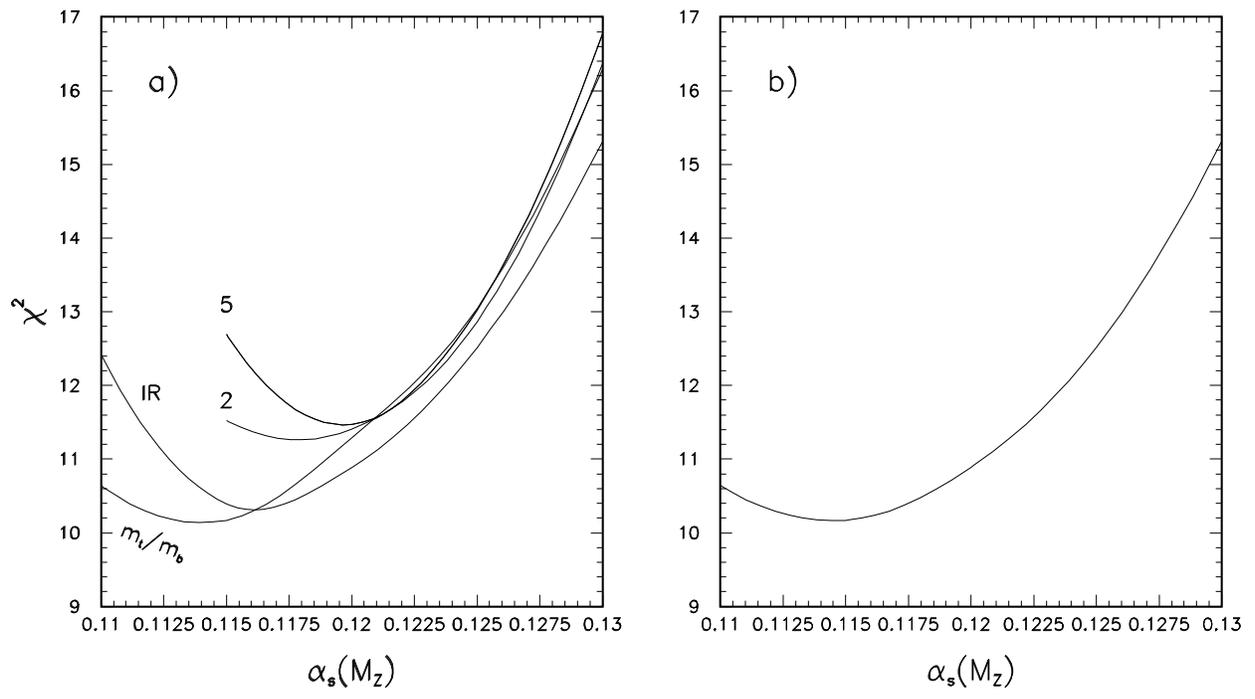
\noindent {\bf Figure 4.}
a) Dependence of ~$\chi^2$ ~on ~$\alpha_s(M_Z)$ ~
for different values of ~$\tan\beta$, ~b)  global dependence
of ~$\chi^2$ ~on ~$\alpha_s(M_Z)$. ~Only version A is shown;
in version B the results are very similar.
\vskip 0.3cm

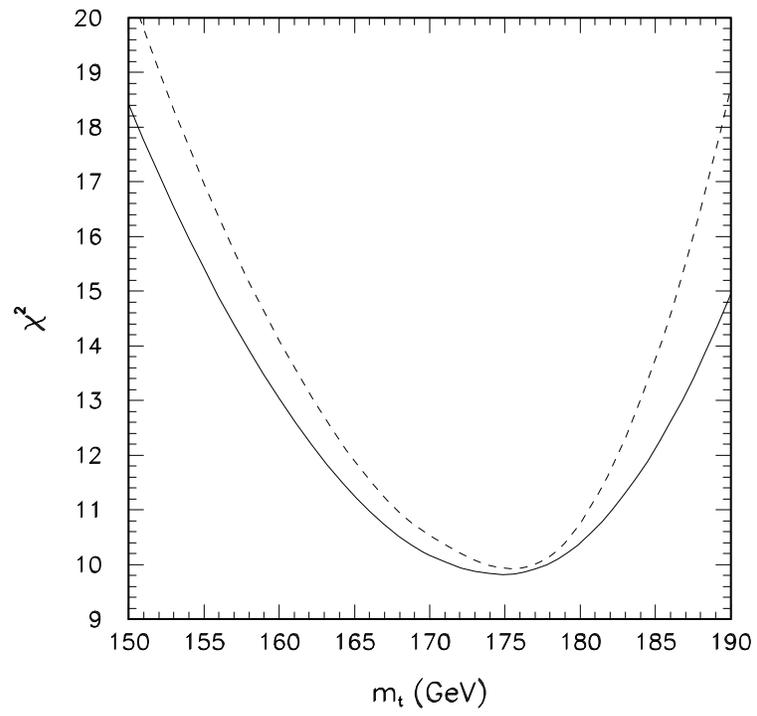
\noindent {\bf Figure 5.}
Global dependence of ~$\chi^2$ ~on ~$m_t$. ~
Solid and dashed lines correspond to versions
A and B of the fit respectively.
\vskip 0.3cm

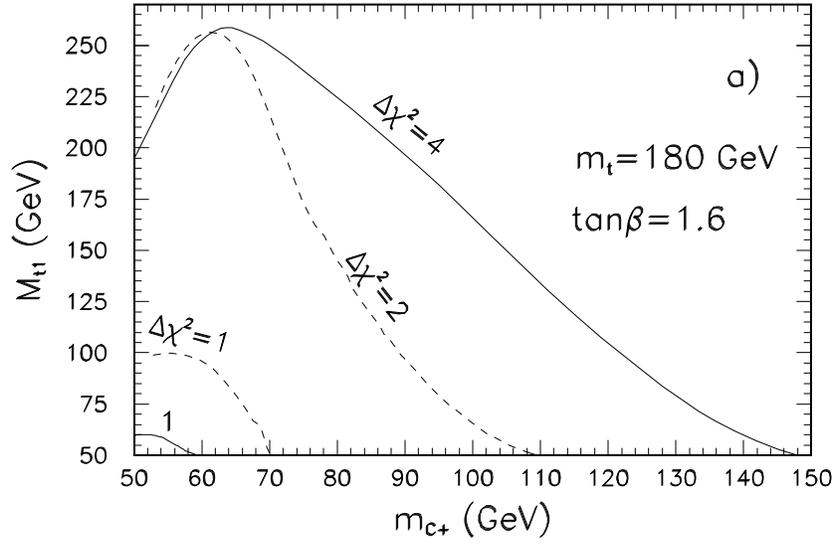
\noindent {\bf Figure 6.}
Contours of constant ~$\Delta\chi^2$ ~plotted in the
chargino -- lighter stop mass plane for low ~$\tan\beta$ ~fits:
a) with ~$\alpha_s(M_Z)$ ~fitted as in version B (solid lines)
and with $\alpha_s(M_Z)$ free as in version A (dashed lines);
b) with ~$\alpha_s(M_Z)$ ~fixed to its best value,
\vskip 0.3cm

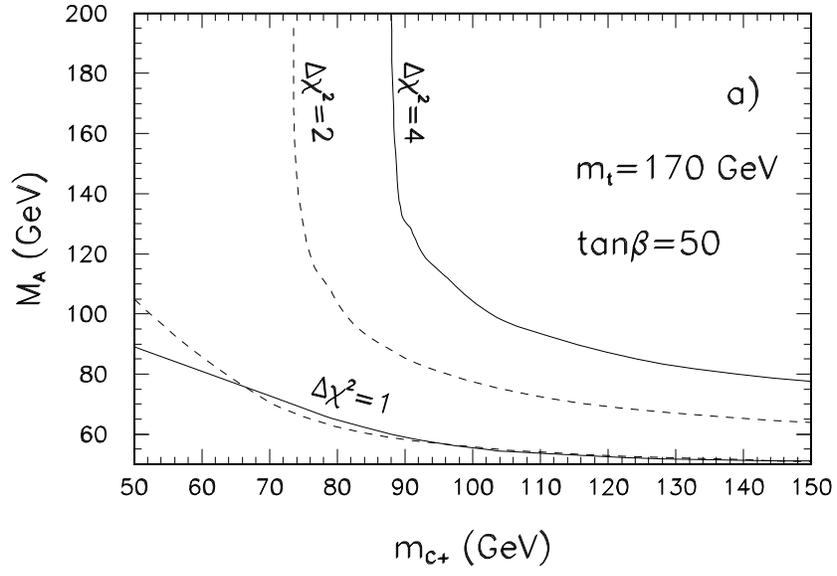
\noindent {\bf Figure 7.}
Contours of constant ~$\Delta\chi^2$ ~plotted in the
chargino -- $CP$--odd Higgs boson mass plane for large ~
$\tan\beta$ ~fits:
a) with ~$\alpha_s(M_Z)$ ~fitted as in version B (solid lines) and with
$\alpha_s(M_Z)$ free as in version A (dashed lines);
b) with ~$\alpha_s(M_Z)$ ~fixed to its best value.
\vskip 0.3cm

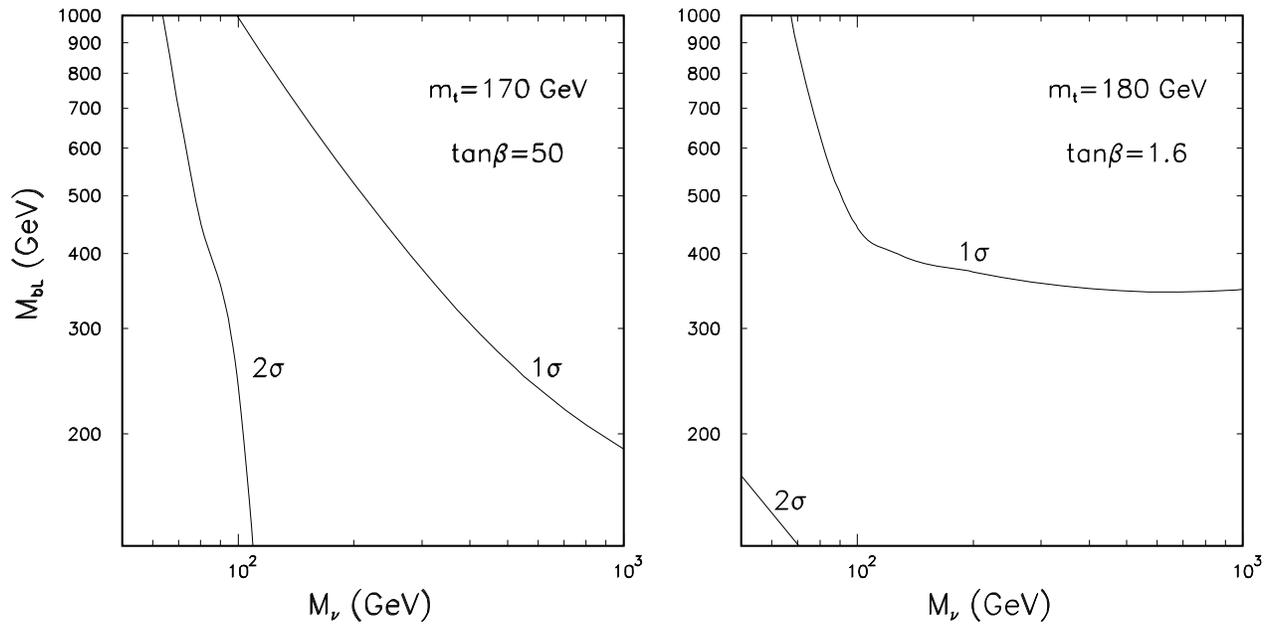
\noindent {\bf Figure 8.}
1 ~and ~2$\sigma$ lower bounds on left--handed sbottom and
slepton masses for different values of ~$m_t$ ~and ~$\tan\beta$. ~
Only version A is shown; in version B the results are very similar.
\vskip 0.3cm

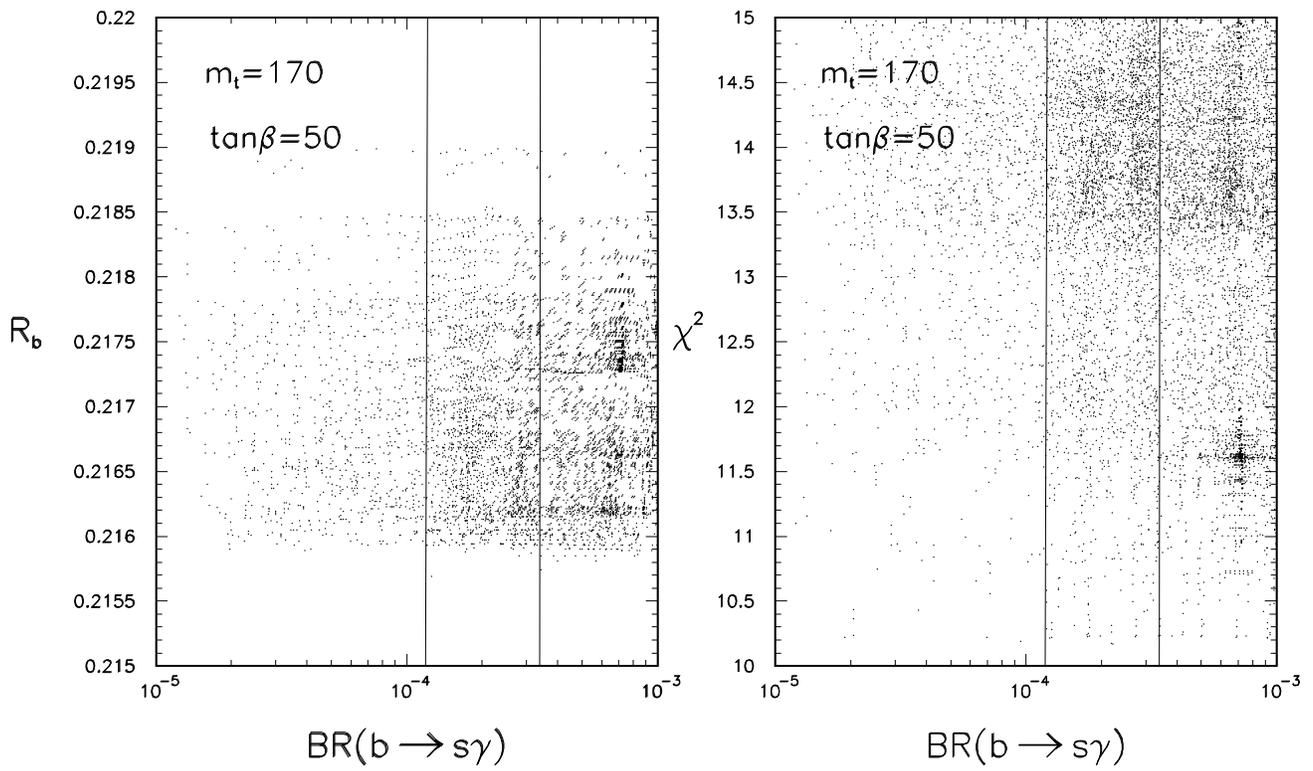
\noindent {\bf Figure 9.}
a) Scatter plot in the plane ~$R_b$, ~$BR(b\rightarrow s\gamma)$ ~
of the points with ~$\Delta\chi^2 < 4$ ~for ~$m_t=170$ GeV ~and
{}~$\tan\beta$=50; ~
b) ~$\chi^2$ ~as a function of ~$BR(b\rightarrow s\gamma)$  for the same values
of ~$m_t$ ~and ~$\tan\beta$.
\vskip 0.3cm

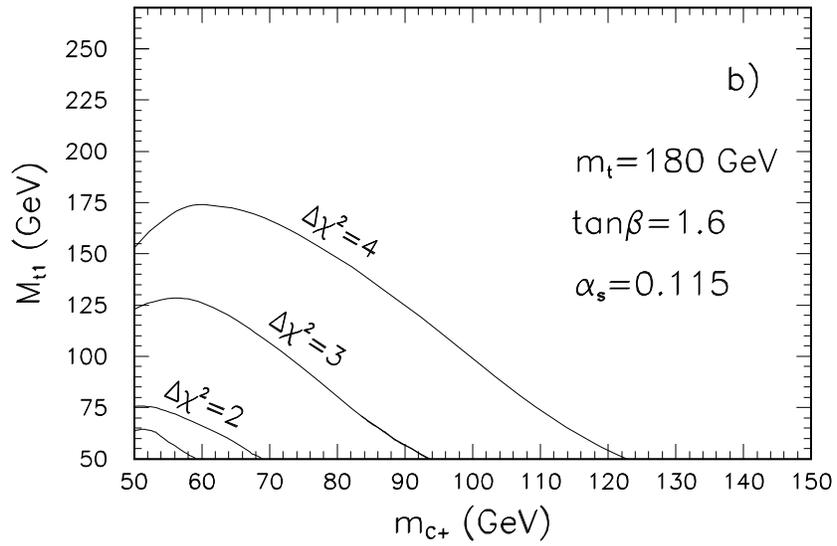
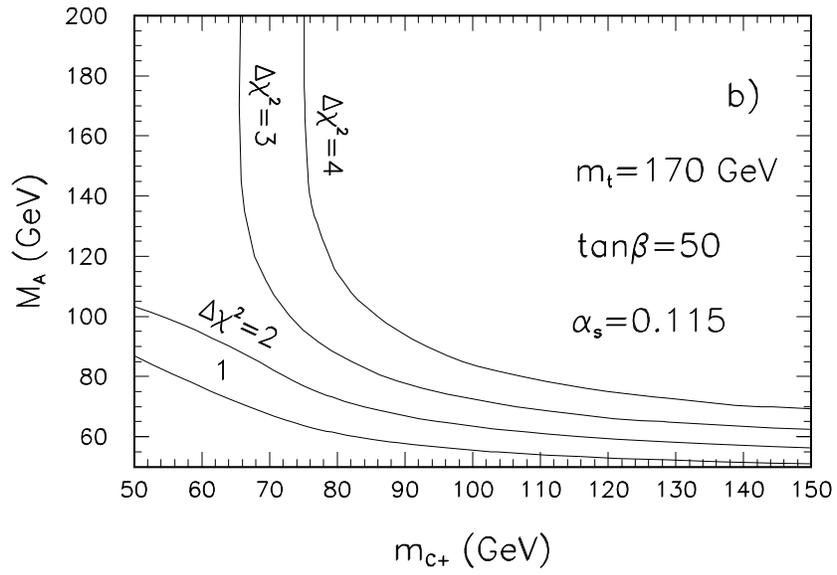
\end{document}